\documentclass[preprint,prd,nofootinbib]{revtex4}
\usepackage[mathscr]{euscript}
\usepackage{bm}
\usepackage{graphicx}
\usepackage{subfigure}
\usepackage{xcolor}
\usepackage[colorlinks=true,linkcolor=red]{hyperref}
\newcommand{\bea}{\begin{eqnarray}}
\newcommand{\eea}{\end{eqnarray}}
\newcommand{\beq}{\begin{equation}}
\newcommand{\eeq}{\end{equation}}
\newcommand{\nn}{\nonumber}
\def\/{\over}

\begin{document}
\title{Quantum fluctuations of spacetime generate quantum entanglement between gravitationally polarizable subsystems}

\author{  Shijing Cheng$^{1}$, Hongwei Yu$^{1}$\footnote{Corresponding author: hwyu@hunnu.edu.cn } and Jiawei Hu$^{1}$\footnote{Corresponding author: jwhu@hunnu.edu.cn}  }

\affiliation{
$^{1}$Department of Physics and Synergetic Innovation Center for Quantum Effects and Applications, Hunan Normal University, Changsha, Hunan 410081, China}

\begin{abstract}

There should  be  quantum vacuum fluctuations of spacetime itself,  if we accept  that  the basic quantum principles we are already familiar with apply as well to a quantum theory of gravity.  In this paper,  we study, in  linearized quantum gravity, the quantum entanglement generation at the neighborhood of the initial time between  two independent gravitationally polarizable two-level subsystems caused by fluctuating quantum vacuum gravitational fields in the framework of open quantum systems.
A bath of fluctuating quantum vacuum gravitational fields serves  as an environment that provides  indirect interactions between the two gravitationally polarizable subsystems, which may lead to entanglement generation.  We find that the entanglement generation is crucially dependent on the polarizations, i.e,  they cannot get entangled in certain circumstances when the polarizations of the subsystems are different while they always can when the polarizations are the same.  We also show that  the presence of a boundary may render  parallel aligned  subsystems   entangled which are otherwise unentangled in a free space.  However, the presence of the boundary does not help
 in terms of entanglement generation if the two subsystems are vertically aligned.

\end{abstract}

\maketitle

\section{Introduction}

Recently, gravitational wave signals from black hole merging systems have been directly detected \cite{Abbotta,Abbotta1,Abbotta2,Abbotta3}, and this confirms the prediction based on Einstein's general relativity over a hundred years ago \cite {gw}. Naturally, one may wonder what happens if gravitational waves are quantized. One of the consequences, if gravity is quantized, is the quantum fluctuations of spacetime itself.  A direct result of spacetime fluctuations is the flight time fluctuations of a probe light signal from its source to a detector~\cite{Ford95,Yua,Yuc}. Another expected effect is the Casimir-like force which arises from the quadrupole moments induced by quantum gravitational vacuum fluctuations \cite{Quach15,Ford15,Wu,Hu,Wu2}, in close analogy to the Casimir and the Casimir-Polder forces \cite{Casimir,cp}.

In the present paper, we are concerned with yet another effect associated with quantum fluctuations of spacetime, that is,  quantum entanglement generation by quantum fluctuations of spacetime.  Quantum entanglement is crucial to our understanding of quantum theory, and it has  many  interesting applications in  various novel technologies. However, a significant challenge to the realization of these quantum technologies with quantum entanglement as a key resource is the environmental noises that lead to the quantum to classical transition.  In the quantum sense, one environment that no
physical system can be isolated from is  the vacuum that fluctuates all the time. On one hand, there are obviously vacuum fluctuations of matter fields, and on the other hand, there should also be  quantum vacuum fluctuations of spacetime itself if we accept  that  the basic quantum principles we are already familiar with apply as well to a quantum theory of gravity.   Currently, there are  discussions with vacuum fluctuations of quantized matter fields as inevitable environmental noises that cause quantum decoherence \cite{decoherence}.

Like matter fields, the fluctuations of gravitational fields, i.e. the fluctuations of spacetime itself, may also cause quantum decoherence. Since  gravitation can not be screened unlike electromagntism, gravitational decoherence is universal. Different models for gravitational decoherence have been proposed \cite{Anastopoulos,Kay,Power,Kok,Wang,Breuer,blhu,Blencowe,Lorenci,Oniga16}, see Ref. \cite{Bassi17} for a recent review. However, decoherence is  not the only role the environment plays. In certain circumstances, the indirect interactions provided by the common bath may also create rather than destroy entanglement between the subsystems via spontaneous emission and photon exchange \cite{Braun,Plenio99,Kim,Schneider,Basharov,Jakobczyk,Reznik,Piani,Floreanini,Zhanga,
Zhangb,Hu15,Yang,Martnez,Valentini,Reznik1,Floreanini1}. Here the bath can be either in the vacuum state  \cite{Jakobczyk,Reznik,Martnez,Valentini,Reznik1,Plenio99,Floreanini1}, or in a thermal state \cite{Braun,Basharov,Schneider,Kim}. In Ref. \cite{Piani}, a general discussion showed that two independent atoms in a common bath can be entangled during a Markovian, completely positive reduced dynamics. In Ref. \cite{Floreanini}, this approach has been applied in the study of the entanglement generation for two atoms immersed in a common bath of massless scalar fields, and found that entanglement generation can be manipulated by varying the bath temperature and the distance between the two atoms. This work was further generalized to the case in the presence of a reflecting boundary \cite{Zhanga,Zhangb}, which showed that the presence of a boundary may offer more freedom in controlling entanglement generation.

In this paper, we study whether a common bath of fluctuating gravitational fields may generate entanglement between subsystems just as matter fields do. We consider a system composed of two gravitationally polarizable two-level subsystems in interaction with a bath of quantum gravitational fields in  vacuum.
We will show that in certain conditions, the two subsystems may get entangled  due to the fluctuating gravitational fields. Like matter fields, the fluctuations of gravitational fields are also expected to be modified if a boundary is present. Although it is generally believed that gravitational waves can hardly be absorbed or reflected,  there have been proposals that the interaction between gravitational waves and quantum fluids might be significantly enhanced compared with normal matter (see Ref. \cite{Kiefer} for a review), and recently there have been interesting conjectures that superconducting films may act as mirrors for gravitational waves due to the so-called Heisenberg-Coulomb effect \cite{hc1,hc2}.  Therefore, we are interested in how the conditions for entanglement generation due to quantum fluctuations of spacetime is modified if such gravitational boundaries exist.

\section{the basic formalism}

The system we consider consists of two gravitationally polarizable two-level subsystems,  which are weakly coupled with a bath of fluctuating quantum gravitational fields in vacuum.
In principle, the gravitationally polarizable system can be any quantum system that carries nonzero quadrupole polarizability. For example, a Bose-Einstein condensate (BEC) is such a system of which the quadrupole polarizability can be estimated~\cite{Hu}.
The Hamiltonian of the whole system takes the following form
\begin{equation}
H=H_{S}+H_{F}+H_{I}.
\end{equation}
Here  $H_{S}$ represents the Hamiltonian of the two gravitationally polarizable two-level subsystems, which can be described as
\begin{equation}
H_{S}=\frac{\omega}{2}\sigma_{3}^{(1)}+\frac{\omega}{2}\sigma_{3}^{(2)},
\end{equation}
where $\sigma_{i}^{(1)}=\sigma_{i}\otimes\sigma_{0}$, $\sigma_{i}^{(2)}=\sigma_{0}\otimes\sigma_{i}$, with $\sigma_{i}\ (i=1,2,3)$ being the Pauli matrices, $\sigma_{0}$ the $2\times2$ unit matrix, and $\omega$ the energy level spacing. $H_{F}$ is the Hamiltonian of the external gravitational fields, the details of which are not needed here. $H_{I}$ describes the quadrupolar interaction between the gravitationally polarizable subsystems and the fluctuating gravitational fields, which can be expressed as
\begin{equation}\label{eq3}
H_{I}=-\frac{1}{2}[Q_{ij}^{(1)}(t)E_{ij}(x^{(1)}(t))+Q_{ij}^{(2)}(t)E_{ij}(x^{(2)}(t))],
\end{equation}
where $Q_{ij}^{(\alpha)}$ $(\alpha=1,2)$ is the induced quadrupole moment of the $\alpha$-th subsystem,  and $E_{ij}=-\nabla_i\nabla_j\phi$  with  $\phi$ being the gravitational potential.
The quadrupolar interaction Hamiltonian~(\ref{eq3}) can be obtained as follows. The energy of a localized mass distribution $\rho_m(x)$ in the presence of an external gravitational potential $\Phi(x)$ is
\begin{equation}\label{v}
V=\int \rho_m(x)\Phi(x)d^3x\;.
\end{equation}
When $\Phi(x)$ varies slowly over the region where the mass is located, it can be expanded as
\begin{equation}\label{phi}
\Phi(x)=\Phi(x_0)+x_i \frac{\partial \Phi(x_0)}{\partial x_i}
    +\frac{1}{2}x_ix_j\frac{\partial^2\Phi(x_0)}{\partial x_i\partial x_j}+\cdots\;,
\end{equation}
so the quadrupolar interaction term reads
\begin{equation}
H_I=\frac{1}{2}\int d^3x \rho_m(x)x_ix_j
   \frac{\partial^2\Phi}{\partial x_i\partial x_j}\;.
\end{equation}
Since $\nabla^2\Phi=0$ in an empty space, the  equation above can be rewritten as
\begin{equation}
H_I=-\frac{1}{2}Q_{ij}E_{ij}\;,
\end{equation}
where
\begin{equation}
Q_{ij}=\int d^3x \rho_m(x)\left(x_ix_j -\frac{1}{3}\delta_{ij}r^2 \right)\;,
\end{equation}
and
\begin{equation}\label{E_ij}
E_{ij}=-\frac{\partial^2\Phi}{\partial x_i\partial x_j}
       +\frac{1}{3}\delta_{ij}\nabla^2\Phi\;.
\end{equation}
In general relativity,  $E_{ij}$ is defined as the Weyl tensor $C_{i0j0}$,  which can be shown to coincide with the expression given in Eq. (\ref{E_ij}) in the Newtonian limit. Here $E_{ij}=C_{i0j0}$ and its dual tensor $B_{ij}=-\frac{1}{2}\epsilon_{ikl}{C^{kl}}_{j0}$ are the gravito-electric and gravito-magnetic tensors which satisfy the linearized Einstein field equations organized in a form similar to the Maxwell equations~\cite{Campbell76,matte53,campbell71,Szekeres,maartens98,Ruggiero02,ramos10,Ingraham}.

In the present paper,
we are considering two quantum subsystems in interaction with quantum gravitational vacuum fluctuations, so the quadrupole moment $Q_{ij}^{(\alpha)}$ ought to be a quantum operator.
In the interaction picture, the quadrupole operator can be written as 
\begin{equation}\label{qij}
Q_{ij}^{(\alpha)}(t)=q_{ij}^{(\alpha)}\sigma_{-}e^{-i\omega t}+q_{ij}^{(\alpha)*}\sigma_{+}e^{i\omega t},
\end{equation}
where {$\sigma_+=
|+\rangle\langle-|$, $\sigma_-=
|-\rangle\langle+|$, are the raising and lowering operators of the subsystem respectively,}
$q_{ij}^{(\alpha)}=\langle{0|Q_{ij}^{(\alpha)}|1}\rangle$ is the transition matrix element of the quadrupole operator of the $\alpha$-th subsystem,
which is  symmetric and traceless, i.e., $q_{ii}^{(\alpha)}=0$ and $q_{ij}^{(\alpha)}=q_{ji}^{(\alpha)}$, thus leaving only five independent components $q^{(\alpha)}_{11}$, $q^{(\alpha)}_{22}$, $q^{(\alpha)}_{12}$, $q^{(\alpha)}_{13}$ and $q^{(\alpha)}_{23}$.
The quadrupole moments are induced by the fluctuating gravitational fields. In this paper, we assume that the gravitational quadrupole polarizabilities are real, and so are the quadrupole moments.
The gravito-electric field $E_{ij}$ is also supposed to be quantized. The metric for a flat spacetime metric with a perturbation can be expressed as $g_{\mu\nu}=\eta_{\mu\nu}+h_{\mu\nu}$, where $\eta_{\mu\nu}$ denotes the flat spacetime metric, and $h_{\mu\nu}$ a linearized perturbation. In the transverse traceless (TT) gauge, the spacetime perturbation can be quantized as  \cite{Yua}
\begin{equation}
h_{ij}=\sum_{\mathbf{k},\lambda}[a_{\mathbf{k},\lambda}e_{ij}(\mathbf{k},\lambda)f_{\mathbf{k}}+H.c.],
\end{equation}
where $f_{\mathbf{k}}=(2\omega(2\pi)^{3})^{-\frac{1}{2}}e^{i(\mathbf{k}\cdot\mathbf{x}-\omega t)}$ is the field mode and $e_{\mu\nu}(\mathbf{k},\lambda)$ is the  polarization tensor with $\omega=|\mathbf{k}|=(k_{x}^{2}+k_{y}^{2}+k_{z}^{2})^{\frac{1}{2}}$. Here H.c. denotes the Hermitian conjugate, $\lambda$ labels the polarization state, and units in which $\hbar=c=32\pi G=1$ have been adopted. From the definition of $E_{ij}$, we have
\begin{equation}\label{E}
E_{ij}=\frac{1}{2}\ddot{h}_{ij},
\end{equation}
where a dot means $\frac{\partial}{\partial t}$.
The  correlation function can then be obtained as~\cite{Yua}
\begin{equation}
{\langle{E_{ij}(x)E_{kl}(x')}\rangle}=\frac{1}{8(2\pi)^{3}}\int d^{3}\mathbf{k}\sum_{\lambda}e_{ij}(\mathbf{k},\lambda)e_{kl}(\mathbf{k},\lambda)\,{\omega^{3}}e^{i\mathbf{k}\cdot(\mathbf{x}-\mathbf{x'})}e^{-i\omega(t-t')},
\end{equation}
where
\bea\label{g2}
&&\sum_{\lambda}e_{ij}(\mathbf{k},\lambda)e_{kl}(\mathbf{k},\lambda)=\delta_{ik}\delta_{jl}+\delta_{il}\delta_{jk}-\delta_{ij}\delta_{kl}+\hat{k_{i}}\hat{k_{j}}\hat{k_{k}}\hat{k_{l}}+\hat{k_{i}}\hat{k_{j}}\delta_{kl}+\hat{k_{k}}\hat{k_{l}}\delta_{ij}\nonumber\\
&&\qquad\qquad\qquad\qquad\qquad\qquad\qquad\qquad-\hat{k_{i}}\hat{k_{l}}\delta_{jk}-\hat{k_{i}}\hat{k_{k}}\delta_{jl}-\hat{k_{j}}\hat{k_{l}}\delta_{ik}-\hat{k_{j}}\hat{k_{k}}\delta_{il},
\eea
with $\hat{k_{i}}={k_{i}}/{k}$.

We assume that initially the two subsystems are decoupled with the quantum gravitational fields, i.e. $\rho_{\rm tot}(0)=\rho(0)\otimes|0\rangle\langle0|$,
where $\rho(0)$  denotes the initial state of the two subsystems, and $|0\rangle$ the vacuum state of the gravitational fields.  The density matrix of the total system  satisfies the Liouville equation
\begin{equation}
\frac{\partial\rho_{\rm tot}(t)}{\partial t} = -i[H,\rho_{\rm tot}(t)].
\end{equation}
Under the Born-Markov approximation, the reduced density matrix of the two gravitationally polarizable subsystems $\rho(t)=\mathrm{Tr}_{F}[\rho_{\rm tot}(t)]$ satisfies the Kossakowskl-Lindblad master equation \cite{Kossakowski,Lindblad},
\begin{equation}\label{m}
\frac{\partial\rho(t)}{\partial t} = -i[H_{\rm eff},\rho(t)]
                                     +\mathcal{D}[\rho(t)],
\end{equation}
where
\begin{equation}
H_{\rm eff}=H_{S}-\frac{i}{2}\sum\limits_{\alpha,\beta=1}^{2}\sum\limits_{i,j=1}^{3}  H_{ij}^{(\alpha\beta)}\sigma_{i}^{(\alpha)}\sigma_{j}^{(\beta)},
\end{equation}
and
\begin{equation}
\mathcal{D}[\rho(t)]=\frac{1}{2}\sum\limits_{\alpha,\beta=1}^{2}\sum\limits_{i,j=1}^{3}  C_{ij}^{(\alpha\beta)}[2\sigma_{j}^{(\beta)}\rho\sigma_{i}^{(\alpha)}-\sigma_{i}^{(\alpha)}\sigma_{j}^{(\beta)}\rho-\rho\sigma_{i}^{(\alpha)}\sigma_{j}^{(\beta)}].
\end{equation}
Introducing the Fourier and Hilbert transforms of the gravitational field correlation function $G_{ijkl}^{(\alpha \beta)}(t-t')=\langle{E_{ij}(t,x_{\alpha})E_{kl}(t',x_{\beta})}\rangle$,
\begin{equation}\label{F}
\mathcal{G}_{ijkl}^{(\alpha\beta)}(\omega)=\int^{\infty}_{-\infty}dt e^{i\omega t}G_{ijkl}^{(\alpha \beta)}(t),
\end{equation}
\begin{equation}
\mathcal{K}_{ijkl}^{(\alpha\beta)}(\omega)=\int_{-\infty}^{\infty}dt\ \mathrm{sgn}{(t)}e^{i\omega t}G_{ijkl}^{(\alpha \beta)}(t)=\frac{P}{\pi i}\int_{-\infty}^{\infty}d\lambda\frac{\mathcal{G}_{ijkl}^{(\alpha\beta)}(\lambda)}{\lambda-\omega},
\end{equation}
where $P$ denotes the principal value.
The coefficient matrix $H_{ij}^{(\alpha\beta)}$ and $C_{ij}^{(\alpha\beta)}$ can then be expressed as
\bea\label{c}
C_{ij}^{(\alpha\beta)}&=&A^{(\alpha\beta)}\delta_{ij}-iB^{(\alpha\beta)}\epsilon_{ijk}\delta_{3k}-A^{(\alpha\beta)}\delta_{3i}\delta_{3j},\\
H_{ij}^{(\alpha\beta)}&=&\mathcal{A}^{(\alpha\beta)}\delta_{ij}-i\mathcal{B}^{(\alpha\beta)}\epsilon_{ijk}\delta_{3k}-\mathcal{A}^{(\alpha\beta)}\delta_{3i}\delta_{3j},
\eea
where
\bea\label{b1}
A^{(\alpha\beta)}&=&\frac{1}{16}[\mathcal{G}^{(\alpha\beta)}(\omega)+\mathcal{G}^{(\alpha\beta)}(-\omega)],\ \
B^{(\alpha\beta)}=\frac{1}{16}[\mathcal{G}^{(\alpha\beta)}(\omega)-\mathcal{G}^{(\alpha\beta)}(-\omega)],\\
\mathcal{A}^{(\alpha\beta)}&=&\frac{1}{16}[\mathcal{K}^{(\alpha\beta)}(\omega)+\mathcal{K}^{(\alpha\beta)}(-\omega)],\ \
\mathcal{B}^{(\alpha\beta)}=\frac{1}{16}[\mathcal{K}^{(\alpha\beta)}(\omega)-\mathcal{K}^{(\alpha\beta)}(-\omega)],
\eea
with
\bea\label{b2}
\mathcal{G}^{(\alpha\beta)}(\omega)&=&\sum\limits_{i,j,k,l=1}^{3}q_{ij}^{(\alpha)*}q_{kl}^{(\beta)}\mathcal{G}_{ijkl}^{(\alpha\beta)}(\omega),\\
\mathcal{K}^{(\alpha\beta)}(\omega)&=&\sum\limits_{i,j,k,l=1}^{3}q_{ij}^{(\alpha)*}q_{kl}^{(\beta)}\mathcal{K}_{ijkl}^{(\alpha\beta)}(\omega).
\eea
Here the effective Hamiltonian $H_{\rm eff}$ corresponds to the Lamb-like shift and an environment induced direct coupling between the two subsystems. In the following, we neglect this term and concentrate on the effects produced by the dissipative part $\mathcal{D}[\rho(t)]$.


\section{The Condition for Entanglement Generation}

In this section, we investigate whether the two subsystems become entangled or not at a neighborhood of the initial time with the help of the partial transposition criterion \cite{Peres,Horodecki}, i.e., a two-atom state $\rho$ is entangled if and only if the operation of the partial transposition of $\rho$ does not preserve its positivity.  We assume that the initial state is $\rho(0)=|+\rangle\langle+|\otimes|-\rangle\langle-|$, which is pure and separable, where $|-\rangle$ and $|+\rangle$ are the ground and excited state respectively. In this case, the partial transposition criterion can be found to be equivalent to the following condition, i.e., entanglement can be generated at the neighbourhood of $t=0$ if and only if \cite{Floreanini,Piani}
\begin{equation}\label{ent-con}
\langle\mu|C^{(11)}|\mu\rangle\langle\nu|(C^{(22)})^{T}|\nu\rangle<|\langle\mu|\mathrm{Re}(C^{(12)})|\nu\rangle|^{2},
\end{equation}
where the subscript $T$ denotes the matrix transposition, and the three-dimensional vectors $\mu_{i}$, $\nu_{i}$ are  $\mu_{i}=\nu_{i}=\{1,-i,0\}$.
For brevity, we make the following substitution in Eq. (\ref{c}), $A^{(11)}\rightarrow A_1$, $A^{(22)}\rightarrow A_2$, and $A^{(12)}, A^{(21)}\rightarrow A_3$, and do the same to the coefficients $B^{(\alpha \beta)}$. Plugging the new expressions of $C_{ij}^{(\alpha \beta)}$ into Eq. (\ref{ent-con}),
leads to
\begin{equation}\label{ab}
A_{3}^{2}+B_{1}B_{2}>A_{1}A_{2}.
\end{equation}
Now we investigate the condition for entanglement generation between  a pair of gravitationally polarizable subsystems both in a free space and in a space with a Dirichlet boundary. In the presence of a boundary, we will consider cases of subsystems aligned parallel to and perpendicular to the boundary respectively.

\subsection{Entanglement generation in the free Minkowski spacetime}

\begin{figure}[!htbp]
\centering\includegraphics[width=0.4\textwidth]{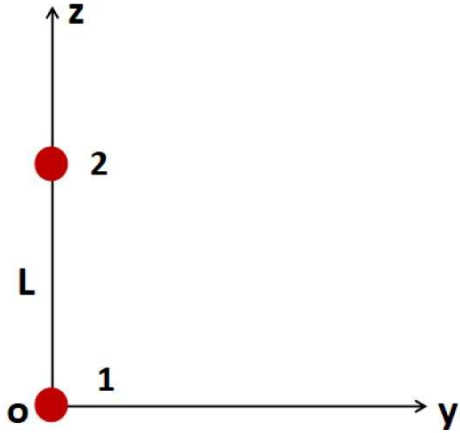}
\caption{\label{F1}
Two gravitationally polarizable subsystems separated from each other by a distance of $L$ in the $z$-axis.}
\end{figure}

We assume the two gravitationally polarizable subsystems are placed at the points ${\bf x}_{1}=(0,0,0)$ and ${\bf x}_{2}=(0,0,L)$ respectively, where $L$ denotes the separation between the two subsystems, as shown in Fig. \ref{F1}.
The explicit expressions of the coefficients $A_i$ and $B_{i}$ in Eq.~(\ref{ab}) are given in appendix \ref{app}, see Eqs. (\ref{A})-(\ref{K}).
Since $A_{1}=B_{1}$ and $A_{2}=B_{2}$, it is clear that the criterion for entanglement generation (\ref{ab}) becomes $A_3^2>0$. That is, entanglement can be generated as long as $A_{3}$, which is dependent on $q_{ij}^{(\alpha)}$ and $\omega L$, is nonzero.

\begin{figure}[!htbp]
\centering\includegraphics[width=0.55\textwidth]{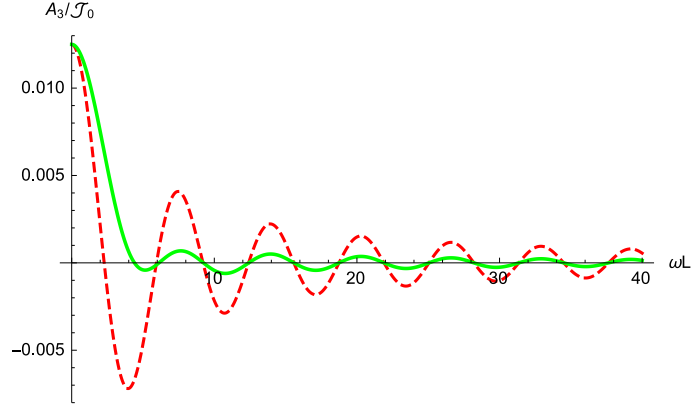}
\caption{\label{wl1}
The coefficient $A_{3}$ as a function of $\omega L$.
The red dashed line and green solid line correspond to two different polarizations of the subsystems respectively, i.e. $q^{(\alpha)}_{11}=-q^{(\alpha)}_{22}=|q|$ and $q^{(\alpha)}_{11}=-q^{(\alpha)}_{33}=|q|$, with other components being zero. }
\end{figure}

When the polarizations of the two subsystems are the same,  for example,  when $q^{(\alpha)}_{11}=-q^{(\alpha)}_{22}=|q|$ or  $q^{(\alpha)}_{11}=-q^{(\alpha)}_{33}=|q|$ and other components being zero, we plot,
in Fig. \ref{wl1},  the coefficient $A_{3}$  as a function of $\omega L$ in the unit of $\Gamma_0$, with $\Gamma_0=|q|\omega^5/\pi$.
As $\omega L$ grows,  $A_{3}$ oscillates around zero with a decreasing amplitude, and it approaches  zero as $\omega L\rightarrow \infty$. That is, when the separation of the two subsystems is infinitely large, they are always separable.
Even for finite  separations, there are some special values of $\omega L$ that give $A_{3}=0$, so entanglement can not be generated at those separations.

When the polarizations of the two subsystems are the same, i.e., $q^{(1)}_{ij}=q^{(2)}_{ij}=q_{ij}$, there are no solutions to $F=K=0$ (see Eqs. (\ref{FF})-(\ref{K})) for any given $L$, i.e., irrespective of  the polarization of the two subsystems, the entanglement between the two subsystems with the same polarization can always be generated for a given separation.
Therefore, when the polarizations of the two subsystems are the same, they can always get entangled if the separation is finite, except for a series of special values of $\omega L$ which are the zero points of $A_3$.
Similar conclusions have been drawn in the case of two independent atoms coupled with massless scalar (matter) fields~\cite{Floreanini}.  This suggests that there is no difference between spacetime fluctuations and matter fields fluctuations when the entanglement generation is concerned with two subsystems of the same gravitational polarization.

When the polarizations of the two subsystems are different, it can be shown by solving the equations $F=K=0$ that the two subsystems may not get entangled for any given separations when the components of quadrupole moments of the two subsystems satisfy the following conditions simultaneously,
\begin{eqnarray}\label{pp}
q_{33}^{(1)}=0, \ \ \ q_{13}^{(1)}\ q_{13}^{(2)}+q_{23}^{(1)}\ q_{23}^{(2)}=0,\ \ \ 2\ q_{12}^{(1)}\ q_{12}^{(2)}+q_{11}^{(1)}\ (q_{11}^{(2)}-q_{22}^{(2)})=0.
\end{eqnarray}
Note that we do not distinguish the two subsystems here, i.e. the superscripts $(1)$ and $(2)$ in the equations above can be exchanged.
Recall that $q_{ij}$ is traceless, so $q_{33}^{(\alpha)}=0$ indicates that $q_{11}^{(\alpha)}=-q_{22}^{(\alpha)}$.
The components of the quadrupole moment can be interpreted as follows. The diagonal components $q_{11}$, $q_{22}$ and $q_{33}$ respectively represent the contributions to the total mass quadrupole moment from the mass distributed along the $x$ axis, $y$ axis and $z$ axis, and the off-diagonal components $q_{12}$, $q_{13}$ and $q_{23}$ respectively represent the contributions to the total mass quadrupole moment from the mass distributed in the $xoy$, $xoz$ and $yoz$ plane.
Therefore, when the mass distributions of the two subsystems induced by quantum gravitational vacuum fluctuations satisfy the condition (\ref{pp}), the two subsystems remain disentangled.

As an example, we consider the case when  the two subsystems are only polarizable  in the $xoy$ plane, i.e. $q^{(1)}_{3i}=q^{(2)}_{3i}=0$. In this case, we have
\bea
{A_{3}}=\frac{1}{128\pi L^{5}}[F_{1}\sin{(\omega L)}+K_{1}\cos{(\omega L)}\ \omega L],
\eea
where
\bea
&&F_{1}=4(3-3\omega^{2}L^{2}+\omega^{4}L^{4})q^{(1)}_{12}q^{(2)}_{12} +(9-5\omega^{2}L^{2}+\omega^{4}L^{4})(q^{(1)}_{11}q^{(2)}_{11}+q^{(1)}_{22}q^{(2)}_{22})\nonumber\\
&&\qquad\qquad\qquad\qquad\qquad\qquad\qquad\ \ \ \ +(\omega^{2}L^{2}-\omega^{4}L^{4}+3)(q^{(1)}_{11}q^{(2)}_{22}+q^{(1)}_{22}q^{(2)}_{11}), \\
\text {and}\nonumber\\
&&K_{1}=4(-3+2\omega^{2}L^{2})q^{(1)}_{12}q^{(2)}_{12}+(2\omega^{2}L^{2}-9)(q^{(1)}_{11}q^{(2)}_{11}+q^{(1)}_{22}q^{(2)}_{22})\nonumber\\
&&\qquad\qquad\qquad\qquad\qquad\qquad\  -(3+2\omega^{2}L^{2})(q^{(1)}_{11}q^{(2)}_{22}+q^{(1)}_{22}q^{(2)}_{11}).
\eea
It is obvious that when
\begin{eqnarray}\label{t}
q_{11}^{(1)}\ q_{11}^{(2)}=q_{22}^{(1)}q_{22}^{(2)}=-q_{12}^{(1)}\ q_{12}^{(2)},
\end{eqnarray}
$A_{3}$ is always zero, i.e., the entanglement generation cannot happen.
Otherwise, the value of ${A_{3}}$ oscillates around zero as $\omega L$ varies with a changing amplitude.
This  reveals a clear difference between entanglement  generation by vacuum  fluctuations of massless scalar (matter) fields and that of the spacetime,  that is, entanglement generation  may never happen for certain polarizations if only spacetime fluctuations are considered.

\subsection{Entanglement generation for two gravitationally polarizable subsystems aligned parallel to the boundary}

\begin{figure}[!htbp]
\centering\includegraphics[width=0.4\textwidth]{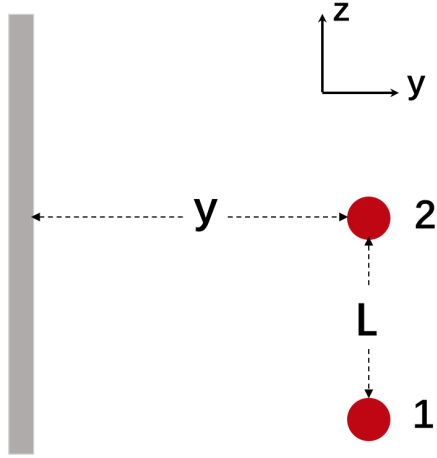}
\caption{\label{F2}
Two independent gravitationally polarizable subsystems are placed along the $z$ direction with a separation $L$, which are at a distance $y$ from a Dirichlet boundary in the $xoz$ plane.}
\end{figure}

Now we concentrate on the effect of a Dirichlet boundary (recall our comments in the Introduction for gravitational boundaries ) on entanglement generation between two gravitationally polarizable subsystems in a bath of fluctuating gravitational fields.  We assume that the Dirichlet boundary is located at $y$=0 in the $xoz$ plane, and the two subsystems separated from each other by a distance $L$ is placed along the $z$ direction at a distance of $y$, see Fig. \ref{F2} .
The coefficients $A_1,~A_2,~A_3$ and $B_1,~B_2,~B_3$ can be calculated from Eq. (\ref{b1}) and Eq. (\ref{b2}) with the Fourier transform of Eq. (\ref{E-parallel}), and the results are lengthy, so we do not give them explicitly here.
As before, we will examine when $A'_{3}=0$ to determine whether entanglement can be generated or not, since $A'_i=B'_i~(i=1,2,3)$ still holds in the present circumstances.
In this case, the conditions that entanglement cannot be generated for any $L$ are solved as follows
\begin{eqnarray}
q_{12}^{(1)}q_{12}^{(2)}=0,\ \ \ q_{13}^{(1)}q_{13}^{(2)}=0,\ \ \ q_{12}^{(1)}q_{13}^{(2)}=0,\ \ \ q_{22}^{(\alpha)}=q_{33}^{(\alpha)}=q_{23}^{(\alpha)}=0\ (\alpha=1\ \mathrm{or}\ 2),
\end{eqnarray}

Now we consider the same example as before, i.e. when the two subsystems are only polarizable  in the $xoy$ plane, $q^{(1)}_{3i}=q^{(2)}_{3i}=0$.  In this case, we have
\bea
A'_{3}=\frac{1}{128\pi}\left[\frac{F_{1}\sin{(\omega L)}+K_{1}\cos{(\omega L)}\ \omega L}{L^{5}}+\frac{M_{1}\sin{(\omega R)}+N_{1}\cos{(\omega R)}\ \omega R}{R^{9}}\right],
\eea
where the explicit expressions of $M_{1}$ and $N_{1}$ are given in appendix (see Eqs. (\ref{m1})-(\ref{n1})).
Obviously, if entanglement generation cannot happen for any given $L$ and $y$, the corresponding coefficients satisfy $F_{1}=K_{1}=M_{1}=N_{1}=0$, and then we have
\begin{eqnarray}\label{tt}
q_{12}^{(1)}q_{12}^{(2)}=0,\ \ \ \ q_{11}^{(1)}q_{11}^{(2)}=0.
\end{eqnarray}
Comparing the result above with that obtained in the corresponding case without a boundary, we find the condition Eq. (\ref{tt}) is a special case of Eq. (\ref{t}).
That is, if the components of the quadrupole moments satisfy $q_{11}^{(1)} q_{11}^{(2)}=q_{22}^{(1)}q_{22}^{(2)}=-q_{12}^{(1)} q_{12}^{(2)}\neq0$, the two subsystems can be entangled in the presence of a boundary, while they remain separable in the free space.

\subsection{Entanglement creation for two gravitationally polarizable subsystems  aligned vertical to the boundary}

\begin{figure}[!htbp]
\centering\includegraphics[width=0.4\textwidth]{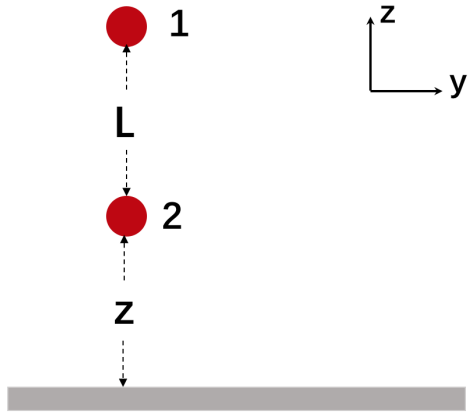}
\caption{\label{F3}
Two independent gravitationally polarizable subsystems are placed on the $z$-axis with a separation $L$. A Dirichlet boundary is located in the $xoy$ plane, and the distance to the closer one is $z$. }
\end{figure}

Now we investigate the case when the alignment of the two gravitationally polarizable subsystems is vertical to the boundary. We assume that the Dirichlet boundary is placed at $z$=0 in the $xoy$ plane and the two subsystems are placed on the $z$-axis with a separation $L$,  the distance from the boundary to the nearer subsystem being $z$, see Fig. \ref{F3}.

Following the same procedures, we find that the conditions that entanglement generation cannot happen for any given $L$ and $z$ are the same with that obtained in the case without a boundary, see Eq. (\ref{pp}). That is, when the quadrupole moments of the two subsystems satisfy certain conditions such that entanglement cannot be generated in the free space, it cannot be generated in the presence of a boundary placed vertically to the alignment of the subsystems either.

\section{Summary}

In this paper, we have investigated the entanglement generation at the neighborhood of the initial time between two independent gravitationally polarizable two-level subsystems in interaction with a bath of fluctuating quantum gravitational fields in vacuum both with and without a boundary. The partial transposition criterion has been applied to determine whether entanglement can be  generated or not at the beginning of evolution. In the free space case, when the polarizations of the two subsystems are the same, 
the two subsystems can always get entangled as long as  the separation is finite. This is similar to what happens when the fluctuations of scalar (matter) fields are considered. When the polarizations of the two subsystems are different, they cannot get entangled  in certain circumstances, whatever the separation is. This is in sharp contrast with the case of massless scalar (matter) fields.  In the presence of a boundary, we find that
in some of the cases where entanglement cannot be generated in a free space, the presence of a boundary placed parallel to the alignment of the subsystems may render the subsystems entangled, but this does not happen for a boundary placed vertically.

\begin{acknowledgments}

This work was supported in part by the NSFC under Grants No. 11435006, No. 11690034, and No. 11805063.

\end{acknowledgments}

\appendix

\section{}\label{app}

The appendix mainly shows some calculations of the coefficients  $A_i$ and $B_i$ in Eq. (\ref{ab}) for the condition of entanglement generation.

\subsection{The case in the free Minkowski spacetime}

In the free Minkowski spacetime, we assume the two gravitationally polarizable subsystems are placed at the points ${\bf x}_{1}=(0,0,0)$ and ${\bf x}_{2}=(0,0,L)$ respectively, where $L$ denotes the separation between the two subsystems.
Straightforward calculations of Eqs. (\ref{g1}) and  (\ref{F}) show that $\mathcal{G}_{ijkl}^{(\alpha\beta)}(-\omega)=0$, and the nonzero components of $\mathcal{G}_{ijkl}^{(\alpha\beta)}(\omega)$ can be written as
\bea
\mathcal{G}_{ijkl}^{(\alpha\beta)}(\omega)=\frac{\omega^{5}}{\pi}f_{ijkl}^{(\alpha\beta)}(\omega, L).
\eea
When $\alpha=\beta$, we have $f_{ijkl}^{(11)}=f_{ijkl}^{(22)}$,
where
\bea\label{f1}
&&f_{1111}^{(11)}=f_{2222}^{(11)}=f_{3333}^{(11)}=\frac{1}{15}, \ \ \ \ \qquad\qquad\qquad f_{1122}^{(11)}=f_{1133}^{(11)}=f_{2233}^{(11)}=-\frac{1}{30},\nonumber \\
&&f_{1212}^{(11)}=f_{1313}^{(11)}=f_{2323}^{(11)}=\frac{1}{20}, \ \ \ \  \qquad\qquad\qquad f_{1122}^{(11)}=f_{2211}^{(11)}=f_{3311}^{(11)}=f_{3322}^{(11)},\nonumber\\
&&f_{1212}^{(11)}=f_{1221}^{(11)}=f_{1331}^{(11)}=f_{2332}^{(11)}=f_{2112}^{(11)}=f_{3113}^{(11)}=f_{3223}^{(11)}=f_{2121}^{(11)}=f_{3131}^{(11)}=f_{3232}^{(11)},
\eea
and when $\alpha\neq\beta$, we have $f_{ijkl}^{(12)}=f_{ijkl}^{(21)}$,
where
\bea
&&f_{1111}^{(12)}(\omega,L)=\frac{\omega L(-9+2\omega^{2}L^{2})\cos{(\omega L)}+(9-5\omega^{2}L^{2}+\omega^{4}L^{4})\sin{(\omega L)}}{8\omega^{5}L^{5}}, \nonumber \\
&&f_{3333}^{(12)}(\omega,L)=\frac{-3\omega L\cos{(\omega L)}+(3-\omega^{2}L^{2})\sin{(\omega L)}}{\omega^{5}L^{5}}, \nonumber\\
&&f_{1122}^{(12)}(\omega,L)=\frac{-\omega L(3+2\omega^{2}L^{2})\cos{(\omega L)}+(3+\omega^{2}L^{2}-\omega^{4}L^{4})\sin{(\omega L)}}{8\omega^{5}L^{5}}, \nonumber\\
&&f_{1133}^{(12)}(\omega,L)=\frac{3\omega L\cos{(\omega L)}+(-3+\omega^{2}L^{2})\sin{(\omega L)}}{2\omega^{5}L^{5}}, \nonumber\\
&&f_{1212}^{(12)}(\omega,L)=\frac{\omega L(-3+2\omega^{2}L^{2})\cos{(\omega L)}+(3-3\omega^{2}L^{2}+\omega^{4}L^{4})\sin{(\omega L)}}{8\omega^{5}L^{5}}, \nonumber\\
&&f_{1313}^{(12)}(\omega,L)=\frac{\omega L(6-\omega^{2}L^{2})\cos{(\omega L)}+3(-2+\omega^{2}L^{2})\sin{(\omega L)}}{4\omega^{5}L^{5}}, \nonumber\\
&&f_{1111}^{(12)}=f_{2222}^{(12)}, \ \ \ \ \ \qquad\qquad\qquad f_{1122}^{(12)}=f_{2211}^{(12)}, \nonumber\\
&&f_{1133}^{(12)}=f_{2233}^{(12)}=f_{3311}^{(12)}=f_{3322}^{(12)}, \ \ \ f_{1212}^{(12)}=f_{1221}^{(12)}=f_{2112}^{(12)}=f_{2121}^{(12)}, \nonumber\\
&&f_{1313}^{(12)}=f_{2323}^{(12)}=f_{1331}^{(12)}=f_{2332}^{(12)}=f_{3113}^{(12)}=f_{3223}^{(12)}=f_{3131}^{(12)}=f_{3232}^{(12)}.
\label{f2}
\eea
The coefficients can be obtained from Eq. (\ref{b1}) and Eq. (\ref{b2}) as
\bea\label{AA1}
A_{1}=B_{1}&=&\frac{\omega^{5}}{480\pi}
\big\{(q^{(1)}_{11}-q^{(1)}_{22})^{2}+(q^{(1)}_{11}-q^{(1)}_{33})^{2}+(q^{(1)}_{22}-q^{(1)}_{33})^{2}\nn \\
&&\qquad\quad\qquad\qquad\quad+6[({q^{(1)}_{12}})^{2}+({q^{(1)}_{13}})^{2}+({q^{(1)}_{23}})^{2}]\big\} \nonumber,
\eea
\bea\label{AA2}
A_{2}=B_{2}&=&\frac{\omega^{5}}{480\pi}
\big\{(q^{(2)}_{11}-q^{(2)}_{22})^{2}+(q^{(2)}_{11}-q^{(2)}_{33})^{2}+(q^{(2)}_{22}-q^{(2)}_{33})^{2}\nn\\
&&\qquad\quad\qquad\qquad\quad+6[({q^{(2)}_{12}})^{2}+({q^{(2)}_{13}})^{2}+({q^{(2)}_{23}})^{2}]\big\} \nonumber,
\eea
\bea\label{A}
A_{3}=B_{3}=\frac{1}{128\pi L^{5}}[F\sin{(\omega L)}+K\cos{(\omega L)}\ \omega L],
\eea
with
\bea\label{FF}
&&F=(9-5\omega^{2}L^{2}+\omega^{4}L^{4})(q^{(1)}_{11}q^{(2)}_{11}+q^{(1)}_{22}q^{(2)}_{22})+(3+\omega^{2}L^{2}-\omega^{4}L^{4})(q^{(1)}_{11}q^{(2)}_{22}+q^{(1)}_{22}q^{(2)}_{11})\nonumber\\
&&\qquad\qquad\qquad\ \ \ \ +4(3-3\omega^{2}L^{2}+\omega^{4}L^{4})q^{(1)}_{12}q^{(2)}_{12} +24(-2+\omega^{2}L^{2})(q^{(1)}_{13}q^{(2)}_{13}+q^{(1)}_{23}q^{(2)}_{23}) \nonumber\\
&&\qquad\qquad\qquad\qquad\ +4(\omega^{2}L^{2}-3)(q^{(1)}_{33}q^{(2)}_{11}+q^{(1)}_{11}q^{(2)}_{33}+q^{(1)}_{33}q^{(2)}_{22}+q^{(1)}_{22}q^{(2)}_{33}-2q^{(1)}_{33}q^{(2)}_{33}),
\eea
\text {and}\nonumber
\bea\label{K}
&&K=(2\omega^{2}L^{2}-9)(q^{(1)}_{11}q^{(2)}_{11}+q^{(1)}_{22}q^{(2)}_{22})-(3+2\omega^{2}L^{2})(q^{(1)}_{11}q^{(2)}_{22}+q^{(1)}_{22}q^{(2)}_{11})\nonumber\\
&&\qquad\qquad\ +4(-3+2\omega^{2}L^{2})q^{(1)}_{12}q^{(2)}_{12}-8(-6+\omega^{2}L^{2})(q^{(1)}_{13}q^{(2)}_{13}+q^{(1)}_{23}q^{(2)}_{23})\nonumber\\
&&\qquad\qquad\qquad\ \ \ +12(q^{(1)}_{33}q^{(2)}_{11}+q^{(1)}_{11}q^{(2)}_{33}+q^{(1)}_{33}q^{(2)}_{22}+q^{(1)}_{22}q^{(2)}_{33}-2q^{(1)}_{33}q^{(2)}_{33}).
\eea

\subsection{The case for two subsystems aligned parallel to the boundary}

As for two gravitationally polarizable subsystems aligned parallel to the boundary, we assume that the boundary is located at $y$=0 in the $xoz$ plane, and the two subsystems separated from each other by a distance $L$ is placed along the $z$ direction at a distance of $y$.
With the help of the method of images, the correlation functions can be written in the following form
\begin{equation}\label{E-parallel}
\langle{E_{ij}(t,x_{\alpha})E_{kl}(t',x_{\beta})}\rangle_{\mathrm{tot}}={\langle{E_{ij}(t,x_{\alpha})E_{kl}(t',x_{\beta})}\rangle}_{\mathrm{free}}-{\langle{E_{ij}(t,x_{\alpha})E_{kl}(t',x_{\beta})}\rangle}_{\mathrm{bnd}},
\end{equation}
and the coefficients $A'_1,~A'_2,~A'_3$ and $B'_1,~B'_2,~B'_3$ can be calculated from Eqs. (\ref{b1})-(\ref{b2}) with the Fourier transform of Eq. (\ref{E-parallel}).
The results are lengthy, so we do not give them explicitly here, but $A'_i=B'_i~(i=1,2,3)$ still holds as it does in the free space.

When the two subsystems are only polarizable in the $xoy$ plane, i.e., $q^{(1)}_{3i}=q^{(2)}_{3i}=0$, we have
\bea
A'_{3}=\frac{1}{128\pi}\left[\frac{F_{1}\sin{(\omega L)}+K_{1}\cos{(\omega L)}\ \omega L}{L^{5}}+\frac{M_{1}\sin{(\omega R)}+N_{1}\cos{(\omega R)}\ \omega R}{R^{9}}\right],\eea
where
\bea\label{m1}
&&M_{1}=R^{2}[L^{2}(3-\omega^{2}R^{2}+\omega^{4}R^{4})+8y^{2}(-6+\omega^{2}R^{2}+\omega^{4}R^{4})](q_{11}^{(1)}q_{22}^{(2)}+q_{22}^{(1)}q_{11}^{(2)})\nonumber \\
&&\qquad\quad\ \ -[9L^{4}-288L^{2}y^{2}+384y^{4}+\omega^{2}(L^{6}-108L^{4}y^{2}-192L^{2}y^{4}+1024y^{6})\nonumber \\
&&\qquad\ \ \ \ \ \ +\omega^{4}(L^{4}+12L^{2}y^{2}+32y^{4})^{2}]q_{22}^{(1)}q_{22}^{(2)}+R^{4}(5\omega^{2}R^{2}-\omega^{4}R^{4}-9)q_{11}^{(1)}q_{11}^{(2)}\nonumber \\
&&\qquad\qquad\qquad\qquad\ \ \ \ -4\omega^{2}R^{4}\left[\omega^{2}L^{4}+32\omega^{2}y^{4}+12y^{2}(-1+\omega^{2}L^{2})\right]q_{12}^{(1)}q_{12}^{(2)},
\eea
and
\bea\label{n1}
&&N_{1}=[6L^{2}y^{2}(\omega^{2}R^{2}-18)-128y^{4}(\omega^{2}R^{2}-3)+L^{4}(9+2\omega^{2}R^{2})]q_{22}^{(1)}q_{22}^{(2)}\nonumber \\
&&\qquad\qquad\qquad\qquad\ +R^{2}[-3L^{2}+8y^{2}(6+\omega^{2}R^{2})](q_{11}^{(1)}q_{22}^{(2)}+q_{22}^{(1)}q_{11}^{(2)})\nonumber \\
&&\qquad\qquad\qquad\qquad\quad\ \ \ -R^{4}(2\omega^{2}R^{2}-9)q_{11}^{(1)}q_{11}^{(2)}-48\omega^{2}y^{2}R^{4}q_{12}^{(1)}q_{12}^{(2)}.
\eea

\end{document}